%% file: main.tex
\documentclass[a4paper]{jpconf}
\usepackage{graphicx}
\usepackage{siunitx}
\usepackage{xcolor}
\newcommand{\inches}{\ensuremath{{}^{\prime\prime~}}}
\usepackage{lineno}
\begin{document}
\title{Status and Plans for the CMS High Granularity Calorimeter Upgrade Project}

\author{Thorben Quast}
\address{CERN \\ Esplanade des Particules 1, 1211 Meyrin, CH}
\ead{thorben.quast@cern.ch}

\begin{abstract}
	The CMS Collaboration is preparing to build replacement endcap calorimeters for the HL-LHC era. 
	The new high-granularity calorimeter (HGCAL) is, as the name implies, a highly-granular sampling calorimeter with approximately six million silicon sensor channels ($\approx$\SI{1.1}{\centi\metre}$^2$ or \SI{0.5}{\centi\metre}$^2$ cells) and about 250 thousand channels of scintillator tiles readout with on-tile silicon photomultipliers. 
	The calorimeter is designed to operate in the harsh radiation environment at the HL-LHC, where the average number of interactions per bunch crossing is expected to exceed 140. 
	Besides measuring energy and position of the energy deposits, the electronics is also designed to measure the time of their arrival with a precision in the order of \SI{50}{\pico\second}. 
	This paper summarises the reasoning and ideas behind the HGCAL, describes the current status of the project, and highlights some of the challenges ahead.
\end{abstract}

\input{content/ch1}
\input{content/ch3}
\input{content/ch2}
\input{content/ch4}
\input{content/ch5}

\section*{References}
\bibliography{bib/bib}

\end{document}

%% file: content/ch1.tex
\section{CMS High Granularity Calorimeter Upgrade}
After its Run 3 in the next three years, the Large Hadron Collider~\cite{evans:2008} at CERN will undergo significant upgrades in order to increase its instantaneous luminosity to approximately five times its design value.
This upgraded High Luminosity LHC (HL-LHC)~\cite{hl-lhc-tdr:2017} will be a powerful instrument for experimental testing of the Standard Model of Particle Physics.

\subsection{Motivation}
With the increased instantaneous luminosity, both the expected number of pile-up interactions per bunch crossing, and the anticipated detector damage due to the boosted irradiation levels will increase.
In fact, the original CMS detector concept~\cite{cms:2008}, and in particular its endcap calorimeters~\cite{ecal-tdr:1997,hcal-tdr:1997}, will not function properly in this harsh environment.
Thus, alongside other upgrades, CMS will replace its calorimeter endcaps (CE) for operation at HL-LHC.
Apart from being suitable for the aforementioned experimental conditions, the new calorimeter must be compliant with CMS' particle-flow-based particle and event reconstruction strategy~\cite{particleflow:2017}.

\subsection{Design}
\begin{figure}[h]
	\centering
	\begin{minipage}{0.35\textwidth}
		\centering
		\includegraphics[width=0.75\textwidth]{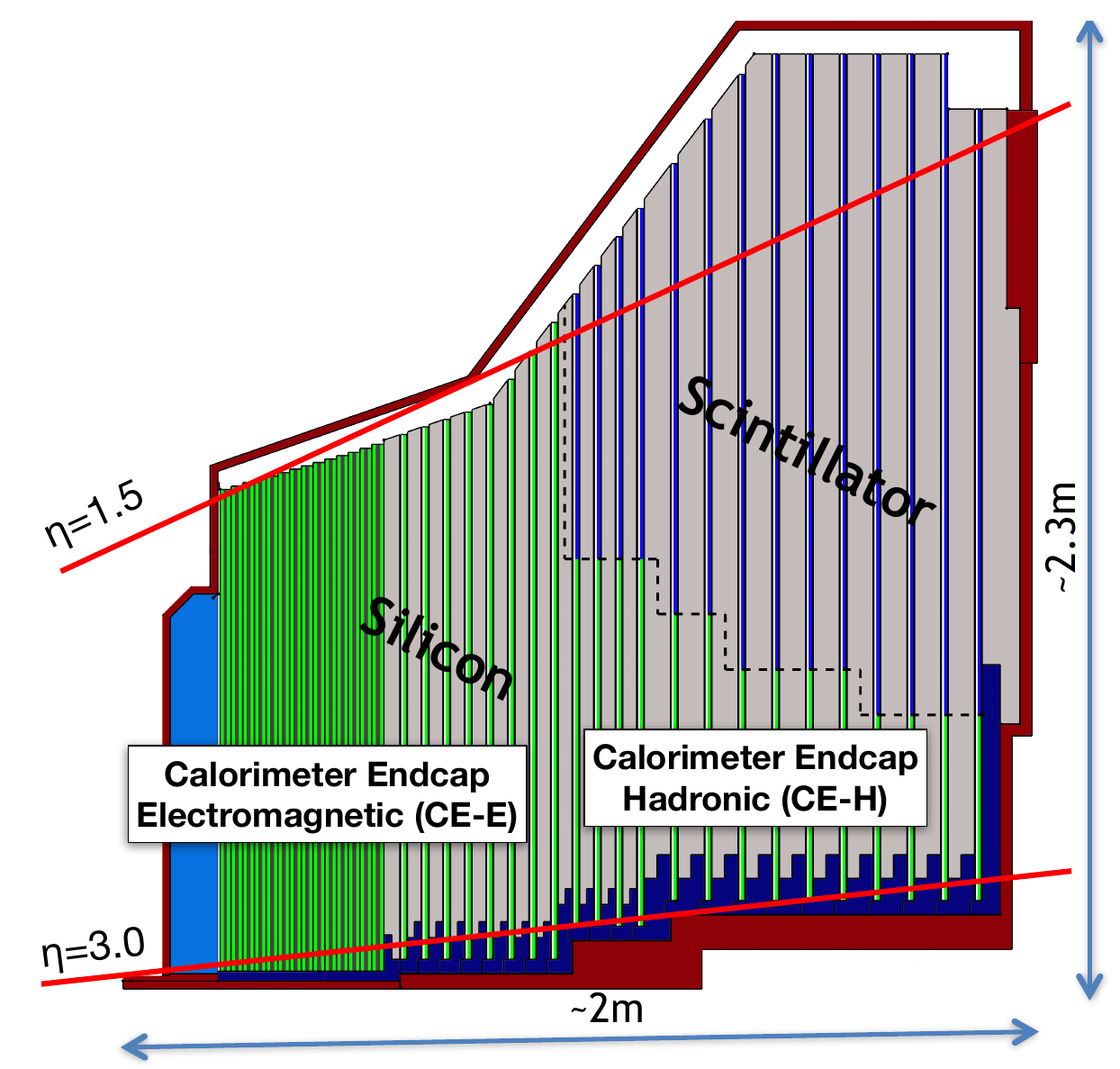}
		\caption{\label{fig:HGC_standard}Cross section of the CMS calorimeter endcap upgrade design.}
	\end{minipage}\hspace{2pc}%
	\begin{minipage}{0.55\textwidth}
		\centering
		\includegraphics[width=0.75\textwidth]{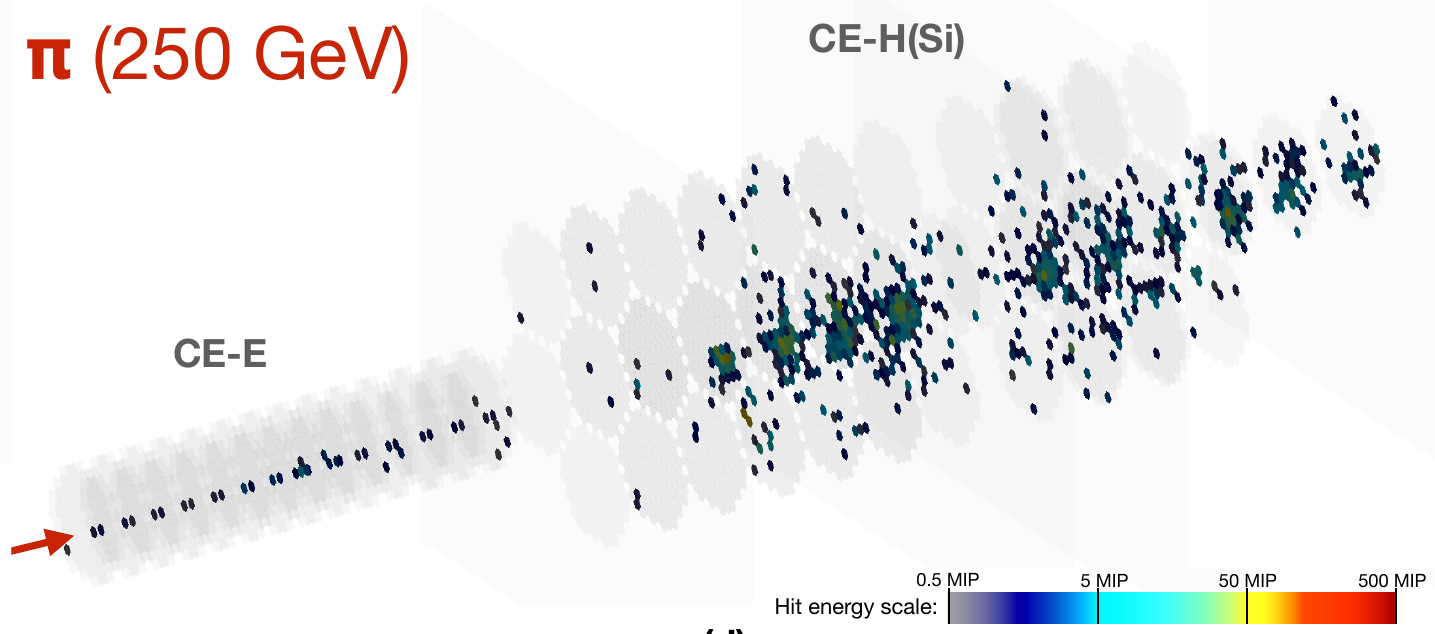}
		\caption{\label{fig:event_displays} A 250$~$GeV charged-pion-induced shower measured in a silicon-based prototype of the new CMS calorimeter endcap.}
	\end{minipage}\hspace{2pc}%
\end{figure}

The new sampling calorimeter is designed to be a highly granular calorimeter (HGCAL). 
In view of their radiation tolerance, compactness, and fast timing capabilities, silicon sensors are used as sensitive material in the forward region of highest expected fluence.
By contrast, the outer regions deploy scintillator-tile and SiPMs-based energy measurements, cf. Figure~\ref{fig:HGC_standard}.
Passive absorbers are primarily made of lead and copper-tungsten in the electromagnetic section (CE-E), and stainless steel in the hadronic section (CE-H).
In order to minimise irradiation-induced sensor degradation, the entire structure will be CO$_2$-cooled down to -30$^{\circ}~$C.
The granularity stems from the readout area of the individual silicon pads of about $~0.5~$-$~1.0~$cm$^2$, cf. Figure~\ref{fig:event_displays}, which allows for spatial disentanglement of energy depositions of overlapping particle showers.

%% file: content/ch3.tex
\section{Challenges}
Core HGCAL components were designed from scratch, prototyped and qualified in the (slowly ending) research and development phase.
Apart from the mechanical engineering aspects, the specifications of the readout electronics, the triggering scheme and the reconstruction of particle showers in the busy HL-LHC environment are particularly challenging.

\subsection{Frontend Electronics}
HGCAL features two different frontend ASICs: The HGCROC~\cite{Thienpont_2020} and the ECON chips.
The signal amplitude is readout by the HGCROC, covering a large dynamic range in this regard to both detect minimum ionising particles (MIPs) and the energy deposits from $\mathcal{O}$(TeV) photon showers.
The upper dynamic range is digitised via a time-over-threshold (TOT) technique whereas the lower range is sampled with a 10$~$bit ADC.
In addition, the HGCROC features a time-of-arrival (TOA) measurement with a 10$~$bit TDC resulting in timing resolutions of a few tens of picoseconds.
There are two output paths, one for the data and one for the trigger, which are forwarded to the ECON-D (data) and ECON-T (trigger) chips, respectively.
The former performs a zero suppression to minimise the required data stream bandwidth.
Both the HGCROC and the ECONs are designed to operate with a limited powering budget of \SI{20}{\milli\watt} per channel, and need to be radiation tolerant.
Their qualification is ongoing.

\subsection{Trigger Implementation}
HGCAL's triggering system will provide three-dimensional clusters of energy deposits to the global CMS L1 trigger~\cite{cms-trigger2017}.
After computing energy sums across neighbouring pads directly in the HGCROC and compression by the ECON-T, trigger cells are formed in a first stage of FPGAs and subsequently are processed further to energy clusters in a second stage of FPGAs.
Hereby, the trigger data bandwidth is effectively reduced by two orders of magnitude.
Machine-learning based autoencoders are currently explored for potential optimisation of the information content at a given bandwidth limitation.

\subsection{Offline Reconstruction}
HGCAL will be the first highly granular calorimeter to operate in a proton-proton collider with sizeable pile-up.
Dedicated offline reconstruction strategies of its data are therefore being developed.
The goal hereby is to first merge energy deposits to two-dimensional clusters on each layer, and then aligning those as graphs representing the actual particle showers.
The underlying strategy is an iterative one~\cite{Pilato_2020}, where electromagnetic signatures are identified and removed from the event first before proceeding with hadronic signatures and subsequently with MIPs.
Due to recent major improvements in its implementation in CMSSW~\cite{cmssw:github}, HGCAL's data reconstruction now takes only 5\% of CMS' computing budget.
In terms of performance, the overall challenge remains the accurate reconstruction of hadronic signatures in high pile-up environments.
Different pattern recognition and machine-learning assisted algorithms are being studied for this purpose.

%% file: content/ch2.tex
\section{Selected Highlights of Recent Progress}
Following the project milestones defined in its technical design report~\cite{hgcal-tdr:2018}, the HGCAL upgrade project has progressed significantly in the last years.
Two concrete examples are shown in the following.

\subsection{Proof-of-Concept in Test Beams 2018}
\begin{figure}[h]
	\centering
	\begin{minipage}{0.80\textwidth}
		\centering
		\includegraphics[width=0.9\textwidth]{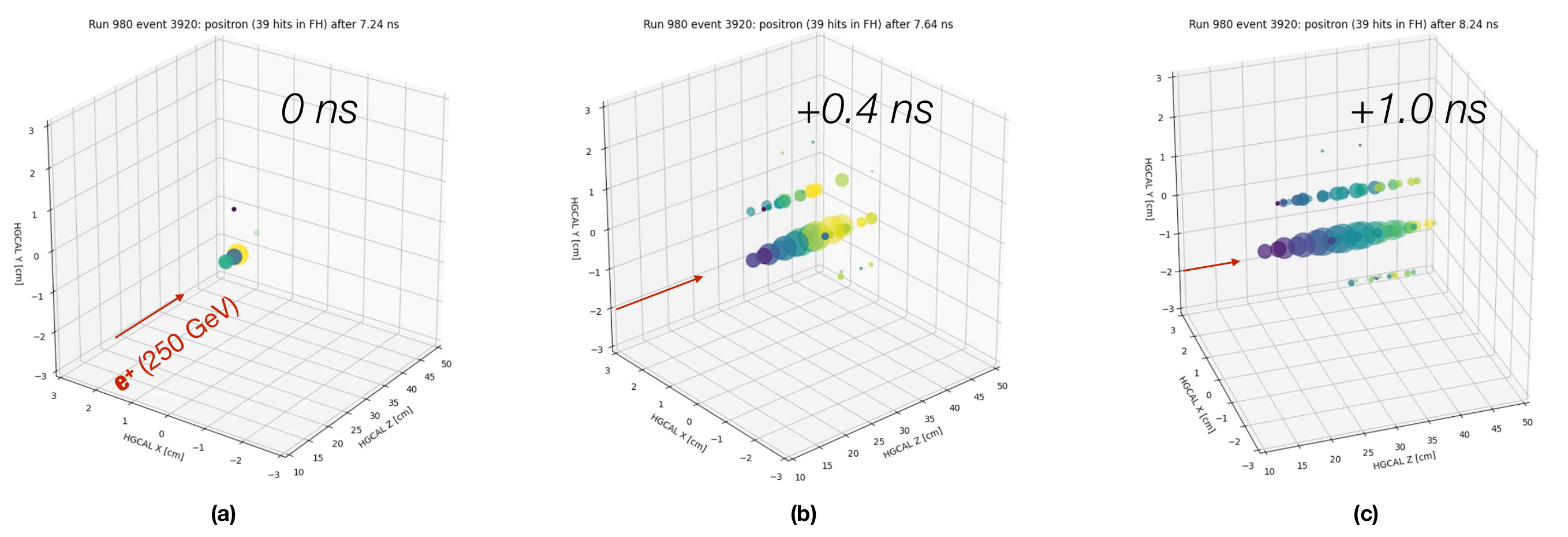}
		\caption{\label{fig:EventVideo} Event "video" of a 250$~$GeV positron-induced shower in a silicon-based prototype of the new CMS calorimeter endcap.}
	\end{minipage}\hspace{2pc}%
\end{figure}

In order to experimentally validate the silicon- based design and the electronics specification parameters, beam tests of HGCAL prototypes have been conducted at the DESY and the CERN SPS test beam facilities in 2018~\cite{H1:2020,H2:2020} using electronics initially developed for the CALICE collaboration~\cite{skiroc2-cms}.
After careful channel inter-calibration of the energy using MIPs, all measured calorimetric performance indicators for electromagnetic and hadronic showers are in agreement with the expectations from GEANT4~\cite{Agostinelli:2002hh}.
Using fast signals from upstream microchannel plates as a reference, time-of-arrival resolutions down to \SI{70}{\pico\second} per readout channel could be observed, which allow to resolve the evolution of particle showers, cf. Figure~\ref{fig:EventVideo}.

\subsection{Qualification of Irradiated Silicon Sensors}
\begin{figure}[h]
	\centering
	\begin{minipage}{0.95\textwidth}
		\centering
		\includegraphics[width=0.9\textwidth]{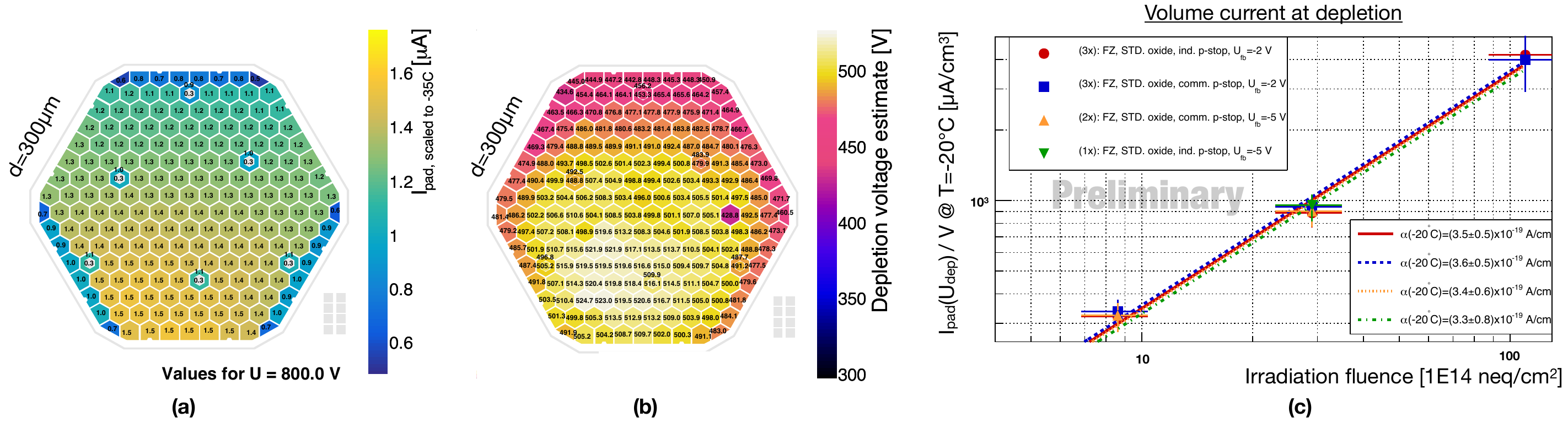}
		\caption{\label{fig:SiSensor} (a) Per-pad leakage current and (b) estimated depletion voltage of a representative \SI{300}{\micro\metre} thick prototype silicon sensor at -35$^\circ~$C after irradiation to $6.5\cdot10^{14}$ 1-MeV-neutron equivalents (neq) per \SI{1}{\centi\metre}$^2$.
		(c) Demonstrated proportionality between the pad leakage current per volume and the fluence for 8\inches prototype HGCAL silicon sensors for different sensor process parameter options under study.
		}
	\end{minipage}\hspace{2pc}%
\end{figure}

As part of the silicon sensor qualification activities, the electrical properties of prototype silicon sensors before and after neutron-irradiation are measured using a custom-made probe- and switchcard system~\cite{pitters:array2019}.
Foremost, the leakage current is a particularly important quantity to understand as it causes power and heat dissipation, and because it introduces higher noise in the system.
In addition, the sensor capacitance as a function of the applied bias voltage allows for assessment of their effective depletion voltage.
Despite Covid-related restrictions in 2020-21, 8\inches prototype silicon sensors could be irradiated at the Rhode Island Nuclear Science Center and subsequently be characterised at CERN at temperatures around HGCAL's foreseen operation temperature.
The preliminary findings are thus far consistent with full functionality of the silicon sensor prototypes up to fluences of $1\cdot10^{16}~$neq/cm$^2$ and demonstrate the expected proportionality between the leakage current per volume and the integrated fluence, cf. Figure~\ref{fig:SiSensor}.

%% file: content/ch4.tex
\section{Getting Ready for Production}
Following the end of the research and development phase, most of the HGCAL components' production is scheduled to start in 2022. 
The corresponding production and assembly sites are currently being prepared and the procurement of the materials has started.
During this process, minor adjustments to the overall HGCAL design are still being made.

\subsection{Silicon Modules Production}
\begin{figure}[h]
	\centering
	\begin{minipage}{0.38\textwidth}
		\centering
		\includegraphics[width=0.51\textwidth]{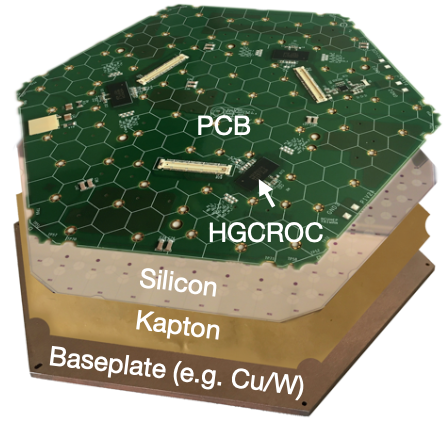}
		\caption{\label{fig:SiModule} Stack up of the silicon-based HGCAL module components.}
	\end{minipage}\hspace{2pc}%
	\begin{minipage}{0.56\textwidth}
		\centering
		\includegraphics[width=0.51\textwidth]{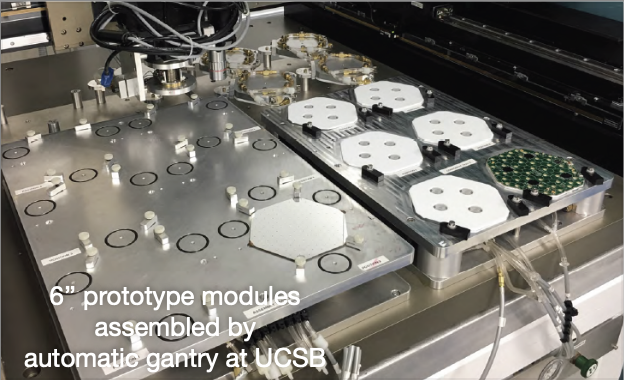}
		\caption{\label{fig:Gantry} Assembly of prototype HGCAL silicon modules by an automatic gantry at UCSB (USA).}
	\end{minipage}\hspace{2pc}%
\end{figure}

The 8\inches silicon-based HGCAL modules consist of glued stacks of a rigid baseplate, gold-plated Kapton\textsuperscript{\texttrademark} foil (for insulation from the baseplate and provision of the bias voltage to the sensor), the silicon sensor and the HGCROC-housing printed circuit board, cf. Figure~\ref{fig:SiModule}.
A total of around 27,000 of such modules will have to be produced for the silicon-based HGCAL compartments.
To minimise the dead space in the silicon-scintillator transition regions and to optimise the inner and outer boundaries, about 20$~\%$ of those modules will have special shapes, i.e. different from full hexagons.
Six different module assembly centres (MACs) are being set up of which five are already fully equipped and are on track to be fully qualified.
With the help of automatic gantries, cf. Figure~\ref{fig:Gantry}, the ultimate goal is assemble more than 10 modules per day in each of the six MACs over a period of approximately two years.

\subsection{Procurement of Passive Absorbers}
The procurement of 600$~$metric tons of stainless steel as CE-H absorbers has been initiated in spring 2021.
Using prototypes of such absorber plates, the feasibility to achieve a flatness of 1$~$mm could be demonstrated.
The corresponding prototyping of the lead-based absorber plates is more challenging due to lead's relative softness and rather low workability, and is still ongoing.

\subsection{Re-Optimisation of the Longitudinal Sampling}
During the mechanical prototyping, it has become evident that the tolerances of the absorber thicknesses might add up to the height required for the placement of the frontend electronics.
Given the necessity to soon proceed with the absorber procurement, the overall HGCAL design has therefore been adapted to tolerances observed on an absorber plate prototype while preserving HGCAL's calorimetric depth.
In particular, the number of sampling layers in the CE-E has been reduced from 28 to 26 and the number of all-silicon-based sampling layers in the CE-H from 8 to 7, respectively.
The benefit of this re-optimisation is the minimisation of the overall risk in HGCAL's construction at the acceptable cost of marginally degrading its expected calorimetric performance, worsening the electromagnetic energy resolution by 4$~\%$.

%% file: content/ch5.tex
\section{Summary}
The CMS High Granularity Calorimeter will replace the existing CMS endcap calorimeters for the high luminosity phase of the Large Hadron Collider.
The upgraded endcap calorimeter will deploy silicon as sensitive material in the high radiation regions, and will consist of scintillator-tiles and SiPM-based energy measurements where the radiation levels permit.
Given the granularity in its readout, three dimensional measurements of shower energies and of the corresponding incidence time will be possible. 
The custom designed frontend electronics, the new triggering system, and the development of the offline reconstruction software are particularly challenging tasks towards the realisation of this ambitious calorimeter.
Recent milestones were the proof-of-principle of the silicon-based design in test beam experiments, and the qualification of prototype silicon sensors. 
Among many other activities, the mass production of modules is being prepared at six different assembly centres, and the procurement of passive absorber plates has been initiated.
In general, the project is currently moving towards a fully-engineered design in order to start mass production in 2022.
Ultimately, the CMS HGCAL will be the first example of highly granular calorimetry used in a high energy frontier running experiment.

%% file: main.bbl
\providecommand{\newblock}{}
\begin{thebibliography}{10}
\expandafter\ifx\csname url\endcsname\relax
  \def\url#1{{\tt #1}}\fi
\expandafter\ifx\csname urlprefix\endcsname\relax\def\urlprefix{URL }\fi
\providecommand{\eprint}[2][]{\url{#2}}

\bibitem{evans:2008}
{L Evans and P Bryant} 2008 {\em JINST\/} {\bf 3} S08001

\bibitem{hl-lhc-tdr:2017}
{B Alonso et al} 2020 {\em {High-Luminosity Large Hadron Collider (HL-LHC):
  Technical design report}\/} (CERN)

\bibitem{cms:2008}
{CMS Collaboration} {2008} {\em {JINST}\/} {\bf 3} S08004

\bibitem{ecal-tdr:1997}
{CMS Collaboration} 1997 {\em {The CMS electromagnetic calorimeter project:
  Technical Design Report}\/}

\bibitem{hcal-tdr:1997}
{CMS Collaboration} 1997 {\em {The CMS hadron calorimeter project: Technical
  Design Report}\/}

\bibitem{particleflow:2017}
{CMS Collaboration} {2017} {\em {JINST}\/} {\bf 12} P10003

\bibitem{Thienpont_2020}
{D Thienpont, C de La Taille} 2020 {\em JINST\/} {\bf 15} C04055

\bibitem{cms-trigger2017}
{CMS Collaboration} 2017 {\em {JINST}\/} {\bf 12} P01020

\bibitem{Pilato_2020}
{A Di Pilato et al} 2020 {\em JINST\/} {\bf 15} C06023

\bibitem{cmssw:github}
{CMS Offline Software} \urlprefix\url{{https://github.com/cms-sw/cmssw}}

\bibitem{hgcal-tdr:2018}
{CMS Collaboration} 2017 {\em {The Phase-2 Upgrade of the CMS Endcap
  Calorimeter}\/} (CERN-LHCC-2017-023)

\bibitem{H1:2020}
{B Acar et al} 2021 {\em JINST\/} {\bf 16} T04001

\bibitem{H2:2020}
{B Acar et al} 2021 {\em JINST\/} {\bf 16} T04002

\bibitem{skiroc2-cms}
{J Borg et al} 2017 {\em {JINST}\/} {\bf 12} C02019

\bibitem{Agostinelli:2002hh}
{S Agostinelli et al} {2003} {\em {NIM A}\/} {\bf {506}}

\bibitem{pitters:array2019}
{E Brondolin et al} 2019 {\em NIM A\/} {\bf 940} 168--173

\end{thebibliography}
